# Multiaxial fields improve SABRE efficiency by preserving hydride order

Shannon L. Eriksson[1,2], Mathew W. Mammen[3], Clark W. Eriksson[4], Jacob R. Lindale[4], Warren S. Warren[5]

[1] Department of Chemistry, Duke University, Durham, NC, 27708

[2] School of Medicine, Duke University, Durham, NC, 27708

[3] Department of Physics, Duke University, NC, 27708

[4] Department of Chemistry, University of Virginia, Charlottesville, VA, 27704

[5] Department of Physics, Chemistry, Biomedical Engineering, and Radiology, Duke University, Durham, NC, 27708

**ABSTRACT**

Signal Amplification By Reversible Exchange (SABRE) and the heteronuclear variant, X-SABRE, increase the sensitivity of magnetic resonance techniques using order derived from reversible binding of para-hydrogen. One current limitations of SABRE is suboptimal polarization transfer over the lifetime of the complex. Here, we demonstrate a multiaxial low-field pulse sequence which allows optimal polarization build-up during a low-field "evolution" pulse, followed by a high-field "mixing" pulse which permits proton decoupling along an orthogonal axis. This preserves the singlet character of the parahydrides while allowing exchange to replenish the ligands on the iridium catalyst. This strategy leads to a 2.5-fold improvement over continuous field SABRE SHEATH experimentally which was confirmed with numerical simulation.

**INTRODUCTION**

Signal Amplification By Reversible Exchange (SABRE) is an inexpensive, generalizable, and quick-acting hyperpolarization method which uses singlet order parahydrogen as the source of spin order. An iridium catalyst transiently binds with both parahydrogen and some target compound in a polarization transfer complex (PTC). In the traditional approach, a small static magnetic field facilitates spin order transfer out of the singlet parahydrides and into spin states with magnetization on the target nucleus. Because the association of these ligands is only transient, chemical exchange occasionally breaks the spin system and reforms in a newly bound PTC.[1,2]

The first implementations of SABRE transferred polarization to $^1$H targets at $mT$ level fields.[1,2] Several variants of SABRE have since been developed since with different targets and field conditions.[3-7] The first experiments to demonstrate efficient, direct pumping of heteronuclei were the LIGHT-SABRE experiments, performed in normal NMR static fields (ca. 10T) where the resonance frequencies are greatly separated, but irradiation of the heteronuclear target with an amplitude $\gamma_1 B_1 \approx J_{HH}$ can be used to establish a resonance condition.[3] Here we focus on heteronuclear SABRE variants (X-SABRE) using low static magnetic fields (mT-µT)[5] which permit polarization transfer to $^{15}$N, $^{31}$P, $^{19}$F, and, notably, $^{13}$C, the predominant target nucleus for clinical imaging applications to date.[8-12] These variants are grounded in the idea of mixing spin eigenstates to transfer population between states at some resonance condition. For example, SABRE SHEATH uses a sub-µT field to make the difference in resonance frequencies for the hydride spins and target heteronuclear spins comparable to the scalar couplings.

An ongoing challenge with SABRE experiments is that the total polarization created tends to be a factor of 5-10 lower than competing (but much more complex and expensive) methods such as dissolution DNP[13]. There is no fundamental reason why this should be true; however, SABRE is much newer than DNP, and the spin dynamics (where exchange rates, Zeeman energy difference, and couplings are all comparable) is complex. More recent experiments have tried to address these issues. One limitation is that probabilistic association and dissociation events interrupt the polarization transfer process at suboptimal times leading to inefficient pumping of polarization on the target nucleus. Coherently-pumped SABRE-SHEATH was introduced to provide experimental control over the polarization transfer dynamics in the setting of exchange.[14] This pulse sequence consists of two pulses. The "evolution" pulse is at the polarization transfer field condition, $B_p$, to allow coherent polarization transfer for the duration of the pulse, $\tau_p$ (typically about 20 ms). The field ($B_d$) is then increased to bring the spin system away from the polarization transfer condition for a much longer duration $\tau_d \approx 200 - 300\ ms$ to allow chemical exchange to "refresh" the singlet parahydrogen and target ligand in the PTC for the next resonant pulse. This coherent pumping can transfer polarization more efficiently by minimizing the randomization effect that exchange can have on the dynamics. However, while the delay field pulse halts singlet order conversion to magnetized spin state, it does not protect the singlet state from interconversion with the triplet manifold because of the symmetry breaking J coupling between the hydrides and target nucleus (or nuclei).[15] Because of this, the singlet order can be partially destroyed during the delay pulse, leading to a decrease in the efficiency of the subsequent polarization transfer pulses.

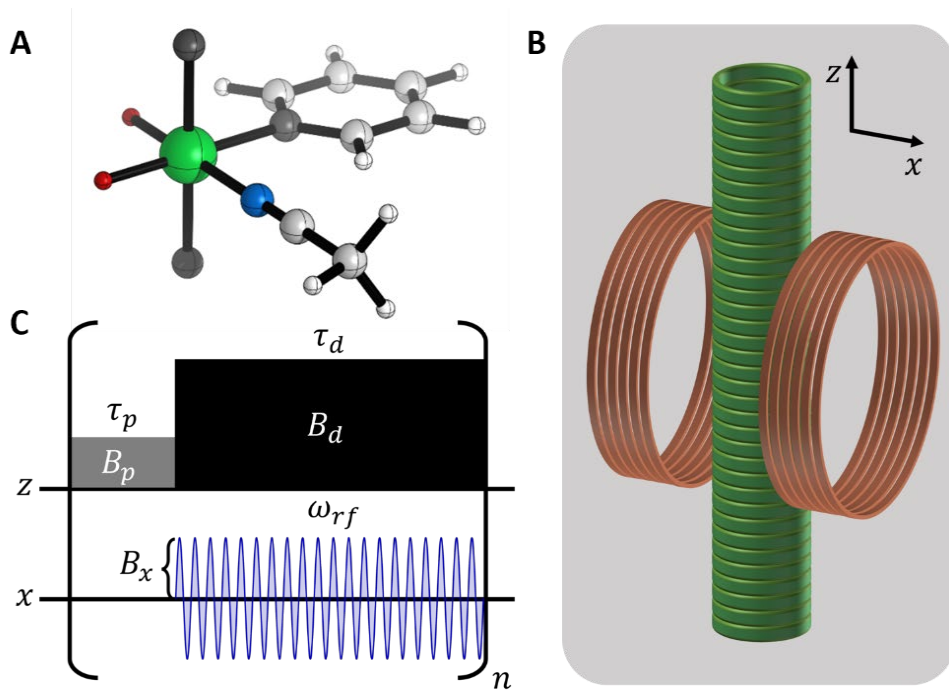

**Figure 1. SABRE hyperpolarization and multiaxial pulse sequence**. (A) SABRE polarization transfer complex with iridium coordination center (green), bound parahydrogen (red), $^{15}$N-acetonitrile target ligand, and $^{14}$N-pyridine co-ligand. (B) Electromagnet array placed in $\mu$-metal shield for field application along the z axis with a piercing solenoid and along the x axis with a Helmholtz coil. (C) Pulse sequence used here. The z pulses are identical to the previously demonstrated coherent SABRE-SHEATH experiment. The x axis sine wave, when resonant with the proton frequency, protects the singlet and increases efficiency.

In a high field regime, the solution to this problem would be obvious: spin decoupling of the hydride resonance[16], which is already in use for many hydrogenative Parahydrogen Induced Polarization (PHIP) experiments.[17,18] The fields used in SABRE and X-SABRE are chosen to make the resonance frequency difference between the hydrides and the target small, making similarly applied selective irradiation problematic. However, these low field experiments have a degree of freedom which is not present in normal NMR: the ability to alter the static field at will. Levitt et. al. have recently introduced a low field multiaxial STORM pulse which excites a population transfer into magnetized target states at low field in PHIP experiments.[19] Here, we introduce a multiaxial low-field pulse sequence which uses a mixing period at high enough field to permit decoupling (Figure 1). We apply a coherently pumped SABRE SHEATH pulse sequence along the z-axis and an oscillating field along a perpendicular axis during the delay period; in a typical application $B_d \approx -50\ \mu T$ and $\omega_{rf} = 2\pi B_d \gamma_H \approx 2\pi(2\ kHz)$. We demonstrate the impact of this oscillating pulse using numerical simulations[20] and then go on to experimentally optimize the parameters of the pulse sequence.

*Coherently Pumped SABRE SHEATH Dynamics*

Let us consider the canonical AA'B SABRE spin system with a PTC with two hydrides, one target nucleus, and one magnetically silent coligand bound. The hydrides are most conveniently written in the singlet-triplet basis set which can be written in the Zeeman basis as

$$|T_{+1,H}\rangle = |\alpha_1\alpha_2\rangle, |T_{0,H}\rangle = \frac{1}{\sqrt{2}}(|\alpha_1\beta_2\rangle + |\beta_1\alpha_2\rangle), |T_{-1,H}\rangle = |\beta_1\beta_2\rangle, |S_H\rangle = \frac{1}{\sqrt{2}}(|\alpha_1\beta_2\rangle - |\beta_1\alpha_2\rangle)$$

Parahydrogen binding overpopulates the $|S_H\rangle$ state of the bound hydrides, but the target nucleus is thermally polarized which can be approximated as 50% $|\alpha_L\rangle$, 50% $|\beta_L\rangle$. This gives us initial states of 50% $|S_H\alpha_L\rangle$ and 50% $|S_H\beta_L\rangle$. SABRE SHEATH then transfers population out of these initialized states into states with opposite signed magnetization on the target nucleus ($|S_H\alpha_L\rangle \to |T_{+1,H}\beta_L\rangle$ and $|S_H\beta_L\rangle \to |T_{-1,H}\alpha_L\rangle$) by evolution under the low-field nuclear spin Hamiltonian:

$$\hat{\mathcal{H}}_{LF} = \omega_H(\hat{I}_{1z} + \hat{I}_{2z}) + \omega_N \hat{L}_z + 2\pi J_{HH}\hat{I}_1 \cdot \hat{I}_2 + 2\pi J_{HL}\hat{I}_1 \cdot \hat{L}$$

Here $\hat{I}_1$ and $\hat{I}_2$ represent the hydrides and $\hat{L}$ represents the target nucleus. The $\omega_H$ and $\omega_L$ terms are the Zeeman interactions of the hydrides and target nucleus, and the $J_{HH}$ and $J_{HL}$ terms represent the couplings between nuclei. Because SABRE SHEATH polarization transfer takes place at low field, all J couplings are in the strong coupling limit. The matrix representation of this Hamiltonian breaks out into 4 orthogonal subspaces detailed in the supplemental information. If the intention is to target the $|S_H\alpha_L\rangle \to |T_{+1,H}\beta_L\rangle$ transition, then the following subspace will pump the desired transition:

$$\begin{array}{c c} & \begin{array}{ccc} |T_{1,H}\beta_L\rangle & |T_{0,H}\alpha_L\rangle & |S_H\alpha_L\rangle \end{array} \\ \begin{array}{c} \langle T_{1,H}\beta_L| \\ \langle T_{0,H}\alpha_L| \\ \langle S_H\alpha_L| \end{array} & \left( \begin{array}{ccc} -\dfrac{\pi J_{HL}}{2} + \omega_H - \omega_L & \dfrac{\pi J_{HL}}{\sqrt{2}} & \dfrac{\pi J_{HL}}{\sqrt{2}} \\ \dfrac{\pi J_{HL}}{\sqrt{2}} & 0 & -\dfrac{\pi J_{HL}}{2} \\ \dfrac{\pi J_{HL}}{\sqrt{2}} & -\dfrac{\pi J_{HL}}{2} & -2\pi J_{HH} \end{array} \right) \end{array}$$

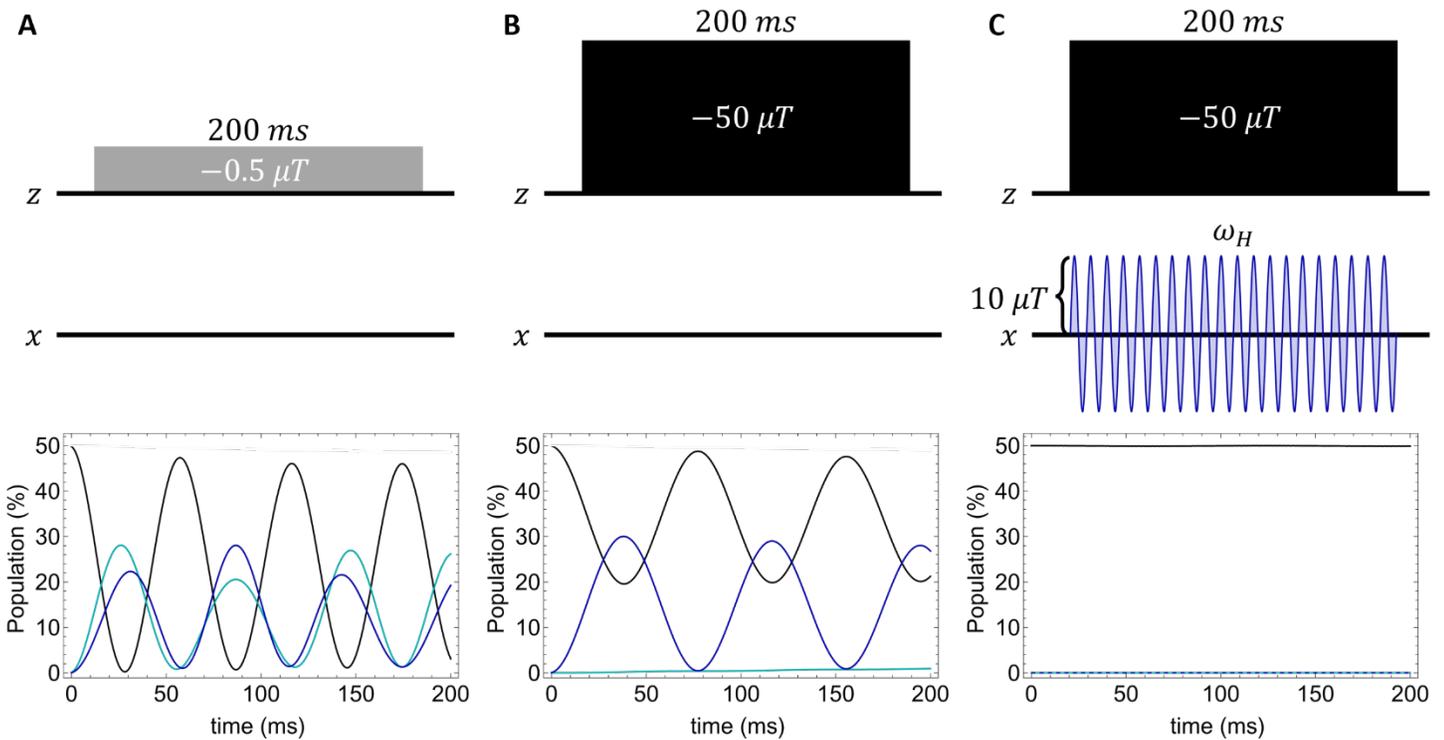

**Figure 2. Evolution of spin populations under various field conditions.** An initial density matrix with 50% population in $|S_H\alpha_N\rangle$ and 50% $|S_H\beta_N\rangle$ is evolved under the pulse conditions specified in the panels and the populations of the initial state, $|S_H\alpha_N\rangle$ (black), target magnetized state, $|T_{1,H}\beta_N\rangle$ (cyan), and off target state, $|T_{0,H}\alpha_N\rangle$ (blue) are shown over time. (A) Evolution under the LAC matching condition shows population transfer out of the initial singlet state into both the target magnetized state and off-target state. (B) Evolution under the "delay" field condition far from resonance, population transfer into the target state is suppressed, but $|S_H\rangle \to |T_{0,H}\rangle$ is allowed by the symmetry breaking J coupling to the target ligand in the complex. (C) Decoupling by irradiating at the proton resonance frequency during the delay pulse preserves the symmetry of the singlet hydrides and prevents population from flowing into the off-target state.

Continuous field SABRE-SHEATH experiments transfer spin population between $|S_H\alpha_L\rangle$ and $|T_{+1,H}\beta_L\rangle$ by bringing the energies of the states together using an externally applied magnetic field. We have recently shown that the Level Anti-Crossing (LAC) matching condition ($B(\gamma_H - \gamma_L) = \frac{\pi}{2}J_{HL} - 2\pi J_{HH}$) is about an order of magnitude off for typical values of $J_{HH}$ and $J_{NH}$.[21]

In coherently pumped SABRE SHEATH the "delay" field brings the spin system far from resonance, halting polarization transfer to allow association of another target ligand and parahydrogen to reinitialize the PTC for the next transfer pulse. During this delay the effective form of the high-field nuclear spin Hamiltonian becomes:

$$\hat{\mathcal{H}}_{HF} = \omega_H(\hat{I}_{1z} + \hat{I}_{2z}) + \omega_L \hat{L}_z + 2\pi J_{HH}\hat{I}_1 \cdot \hat{I}_2 + 2\pi J_{HL}\hat{I}_{1z}\hat{L}_z$$

$$\begin{array}{c|ccc} & |T_{1,H}\beta_L\rangle & |T_{0,H}\alpha_L\rangle & |S_H\alpha_L\rangle \\ \hline \langle T_{1,H}\beta_L| & -\frac{\pi J_{NH}}{2} + \omega_H - \omega_L & 0 & 0 \\ \langle T_{0,H}\alpha_L| & 0 & 0 & -\frac{\pi J_{HL}}{2} \\ \langle S_H\alpha_L| & 0 & -\frac{\pi J_{HL}}{2} & -2\pi J_{HH} \end{array}$$

The initial spin state and target spin state are no longer coupled in the subspace of interest, but singlet parahydrogen is still lost to $|T_{0,H}\rangle$, reducing the polarization transfer efficiency of the following transfer field pulse. Applying a transverse decoupling pulse matching the Larmor frequency of the hydride or target $^{15}N$ nucleus effectively eliminates the heteronuclear coupling during the delay period. Figure 2 shows that any population transfer out of the initialized $|S_H\alpha_L\rangle$ state is prevented under these field conditions.

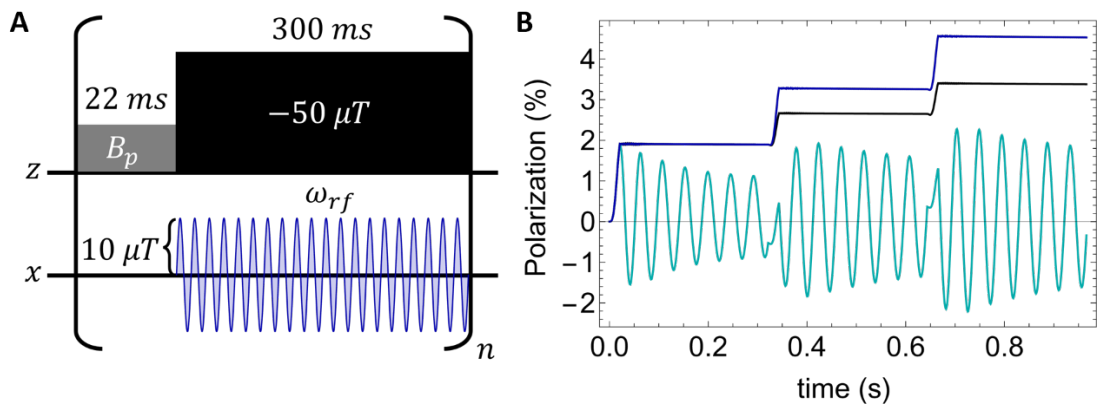

**Figure 3. Polarization build-up for coherently-pumped SABRE SHEATH with and without decoupling.** (A) Pulse sequence used to calculate polarization traces in B. (B) Polarization build-up over time with coherently pumped SABRE SHEATH without delay decoupling in black, with decoupling at $\omega_N$ in cyan, and with decoupling at $\omega_H$ in blue. N magnetization is crushed with irradiation at $\omega_N$, but decoupling on the hydride frequency improves simulated per pulse efficiency.

Using our recently introduced numerical model for exchanging quantum dynamical systems[20], the modified pulse sequence was tested on an arbitrary exchanging 3-spin system. Without decoupling to prevent destruction of $|S_H\rangle$ during the delay pulses the efficiency of subsequent polarization transfer pulses was reduced. Adding a decoupling pulse along a transverse axis preserves the $|S_H\rangle$ population and increases the efficiency of each subsequent transfer pulse (Figure 3). While irradiating on either nuclei would be sufficient to preserve singlet parahydrides, irradiation of the $^{15}N$ nucleus scrambles polarization generated by the "evolution" field pulse. Irradiation on the proton resonance preserves the singlet during the delay without crushing the polarization generated by the pulse sequence to improve per pulse efficiency.

## METHODS

A standardized solution of 2.5 mM IrIMes(COD)Cl [IMes = 1,3-bis(2,4,6-trimethylphenyl)-imidazol-2-ylidene, COD = 1,5-cyclooctadiene], 25 mM natural abundance pyridine, and 50 mM $^{15}$N-acetonitrile was prepared in methanol-$d_4$ and transferred into a high-pressure valved NMR tube. The solution was then bubbled with 43% parahydrogen gas to stimulate the formation of the active form of the PTC (Ir(H)$_2$(IMes)(pyr)$_2$($^{15}$N-acetonitrile)). After activation, SABRE field sequences were applied while bubbling with parahydrogen-enriched gas for a total duration of 60s. The bubbling rate was kept low to minimize evaporation of methanol and concentration of the sample over the course of many experiments.

SABRE SHEATH hyperpolarization requires $\mu T$ level fields for polarization transfer. To achieve this field, a triple $\mu$-metal shield is used to reduce the ambient magnetic field close to 0 $\mu T$. A compensating solenoid is used for fine control over the applied field strength. An additional Helmholtz coil was added for applications of fields perpendicular to the compensating solenoid. The electromagnet array is then driven with a multichannel arbitrary waveform generator to apply a coherently pumped SABRE SHEATH pulse sequence through the solenoid and an oscillating pulse at some frequency $\omega_{rf}$ and amplitude $B_x$ through the Helmholtz coils during the delay field pulse. After 60s under a given set of pulse conditions, the sample was transferred out of the shield and into a 1T $^{15}$N Magritek for signal detection.

## RESULTS AND DISCUSSION

There are six parameters in this pulse sequence to optimize: $B_d$, $B_p$, $B_x$, $\omega_{rf}$, $\tau_p$, and $\tau_d$. The magnitude of the delay field $B_d$ must be large enough to bring the $J_{NH}$ coupling into the weak coupling limit ($\Delta\omega_{NH} \gg J_{NH}$). $B_p$ is set to the optimum continuous field polarization transfer conditions determined by a measure of the enhancement as a function of the applied field. Figure 4 uses an optimized $\tau_p$, arbitrary $\tau_d = 300\ ms$ and sufficiently large $B_x$ to achieve decoupling and the frequency of the oscillating x-field pulse ($\omega_{rf}$) was swept from $30\ Hz$ to $2400\ Hz$. We see three interesting deviations from baseline in these experiments. Near the Larmor frequency of [15]N, there is a steep drop in the enhancement from these pulse sequences. As we are irradiating on resonance with the target nucleus, the delay field of this pulse sequence crushes any polarization built up during each transfer field pulse and the reduction in the final polarization is intuitive. Most importantly, near the Larmor frequency of [1]H, there is an improvement in the final polarization. At this frequency, we are preserving singlet character on the hydrides by preventing the symmetry break caused by coupling out to an external nucleus through the $J_{NH}$ coupling. Because this is done without irradiating on the target nucleus, polarization build-up on [15]N is preserved and amplified by the increase in pulse efficiency. Under the pulse sequence parameters $B_d = -33\ \mu T$ and $B_x = 9\ \mu T$, polarization transfer is improved by 2.5-fold over the continuous field experiment on the same sample ($\epsilon = 3290$ at 1T).

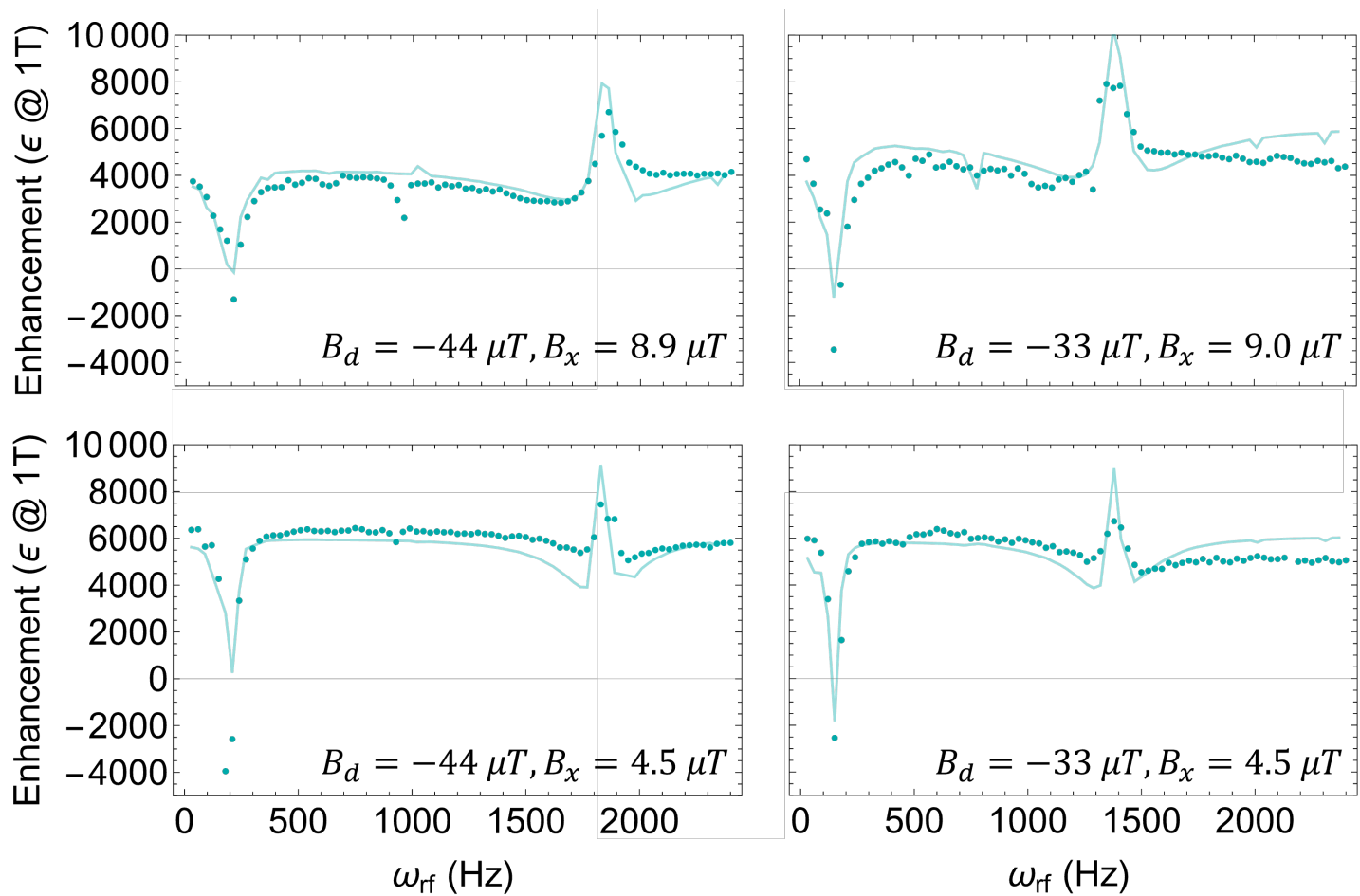

**Figure 4. Experimental enhancement using coherently pumped SABRE SHEATH experiments with delay decoupling at a frequency $\omega_{rf}$.** Experimental enhancements (points) measured after 60s of exposure to the specified field conditions overlaying numerical simulations of the same conditions (lines). The decoupling pulse frequency shows two resonance conditions. At the nitrogen resonance, polarization build-up is suppressed. At the proton resonance, the polarization transfer is improved.

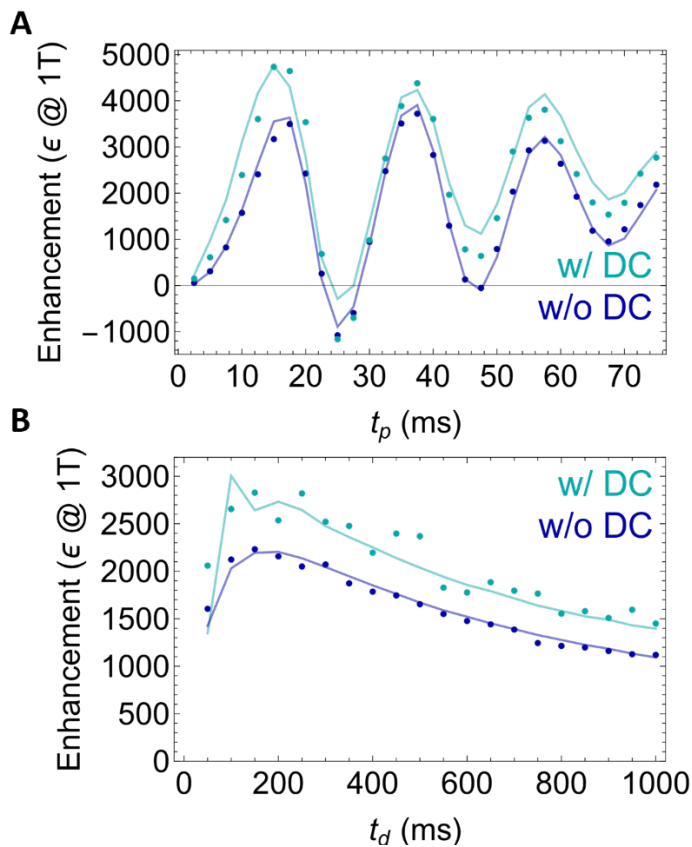

**Figure 5. Experimental and simulated final polarization for coherently pumped SABRE SHEATH experiments with variable pulse durations.** (A) Signal enhancement as a function of $\tau_p$ with and without decoupling at the proton resonance frequency ($\tau_d = 300\ ms$, $k_{d,N} = 12\ s^{-1}$). (B) Signal enhancement as a function of $\tau_d$ with and without decoupling at the proton resonance frequency ($\tau_p = 17.5\ ms$, $k_{a,H} = 0.3\ s^{-1}$). Parameters: $[Ir]:[S] = 1:20$, $B_P = -0.88\ \mu T$, $B_d = -44\ \mu T$, $\omega_{rf} = 1860\ Hz$, $B_x = 4.5\ \mu T$.

Finally, there is a dip in polarization at about half the proton resonance frequency. This may have a contribution from sine wave imperfections, but it is seen in simulations as well as a manifestation of two-photon absorption in low-field NMR[22], made possible by the large magnitude of the transverse field relative to the leading field. When this experiment was performed at different delay fields, these features moved proportionally to the changed field condition. The amplitude of irradiation must be sufficiently large to decouple sufficiently, but otherwise makes no improvements at higher fields. See the supplemental information for additional detail.

We can now reexamine the optimal coherently pumped experiment both with and without decoupling. Using the same field conditions for polarization transfer and delay, we optimized for the pulse durations. The quantum dynamical oscillations which dominate the dependence on $\tau_p$ will be unaffected by the decoupling during the delay pulse. However, as the decoupling improves the "reinitialization" effect of the delay pulse, the range of optimal delay field durations, $\tau_d$, is slightly longer. The improvement in per pulse efficiency at longer delay durations better offsets the effects of relaxation. With full optimization of all parameters we have achieved an enhancement of 2.5-fold over the optimal continuous field experiment on the same solution.

## CONCLUSION

Here we demonstrate that using an oscillating transverse field in in combination with an off-resonant field condition effectively eliminates the heteronuclear $J_{NH}$ coupling to prevent symmetry breaking of the singlet state in SABRE PTCs. Both numerical simulation and experiment demonstrate a clear improvement in polarization transfer using this deocupling delay to preserve singlet order for subsequent evolution field pulses. After optimization of pulse parameters, up to a 3.5-

fold improvement over continuous field SABRE-SHEATH was found. While coherently pumped SABRE SHEATH provided an obvious application for low-field decoupling in SABRE SHEATH, other potential applications include preservation of parahydrogen singlet order in samples with a fixed amount of parahydrogen or controlling the initiation of polarization transfer out of the singlet state to more accurately measure the quantum dynamics of the spin system of interest.

# Acknowledgments

**Funding:**
This work was supported by the National Science Foundation under grant CHE-2003109.

**Author contributions:** CRediT

    Conceptualization: SLE, WSW, JRL, MWM, CWE
    Methodology: SLE, CWE
    Software: SLE, MWM, JRL
    Formal Analysis: SLE, MWM, JRL
    Investigation: SLE, MWM
    Visualization: SLE, WSW
    Supervision: WSW
    Writing—original draft: WSW, SLE
    Writing—review & editing: WSW, SLE

**Competing interests:** Authors declare that they have no competing interests.

**SUPPLEMENTAL INFORMATION**

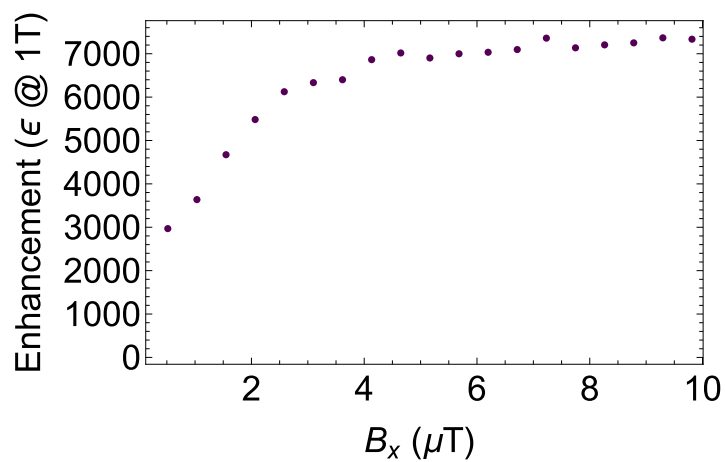

**Figure S1. $B_x$ amplitude dependence on signal enhancement with decoupling pulse.** Experimental signal enhancement from coherently pumped SABRE SHEATH with oscillating x-field at variable amplitude oscillations. Parameters: $[Ir]:[S] = 1:20$, $B_P = -0.88\ \mu T$, $B_d = -44\ \mu T$, $\omega_{rf} = 1860\ Hz$, $\tau_p = 18\ \mu T$, $\tau_d = 300\ ms$.